# A quantum computing scheme for the Hamiltonian path problem

G. Sampath
Department of Computer Science, The College of New Jersey, Ewing, NJ 08628

A quantum computing scheme that uses a single photon and multiple-slit gratings is suggested for the Hamiltonian path problem on a simple graph G of N vertices. The photon is input to an N-slit grating that is followed by an N × N matrix of 'processing units'. A unit consists of a delay line with a delay $\delta_j$ that is distinct for each column j followed by a grating with $k_j$ slits ($1 \leq k_j < N$) whose outputs are directed to $k_j$ units in the next row. There is a one-to-one mapping between paths of length N-1 in the graph and physical paths through the matrix. At the quantum mechanical level the photon's path is a superposition of all these physical paths. The time taken by the photon along a physical path corresponding to a Hamiltonian path in G is a fixed value equal to the sum of the N distinct $\delta_j$'s, and is different from the time along any other path. The graph is Hamiltonian if any one of N detectors placed in the output of the N units in row N detects the photon at this fixed time.

## 1. Introduction

A Hamiltonian path in a simple graph G of N vertices is a sequence of N vertices such that the N-1 successive vertex pairs in the sequence correspond to edges in the graph and each of the N vertices occurs exactly once [1]. Determining whether G has such a path is known to be NP-complete for any known algorithm based on the classical Turing model [2] in both the deterministic and probabilistic versions. More recently quantum mechanical versions of the Turing machine (QTM) [3] have been studied and shown to have more computing power than a probabilistic Turing machine. Quantum algorithms that are more efficient than any known classical algorithm are available for two non-trivial problems: factoring of an integer [4] and search of an unsorted database [5]. Whether a QTM is capable of solving NP-complete problems in polynomial time is not known to date, although it has recently been shown that small non-linearities in the evolution of quantum states can be used to solve not only NP-complete but also #P-complete problems in polynomial time [6]. (Quantum computing theory by and large is based on the assumption that quantum mechanics is linear.)

In the present work, a somewhat different approach is taken. A quantum computing scheme that uses a system of delays to detect a Hamiltonian path in a simple graph is suggested. It is adapted from a recently proposed model of parallel computing that operates in the classical domain. In that model, described in [7], a single electronic or optical pulse is input to a feed-forward delay network, with copies of the pulse generated at each level of the network in a manner determined by the adjacency properties of the graph. The pulses so generated are delayed by preset amounts (N distinct delay values are used) as they travel through the network and are detected at the output of the network. The paths taken by them map one-to-one to paths of length N-1 in the graph. For a suitable choice of delay values the graph has a Hamiltonian path if and only if a pulse is detected at the output at time equal to the sum of the N distinct delays. This arrangement is essentially an interference scheme that behaves like a nondeterministic Turing machine, its behavior is understood in terms of classical physics.

In contrast, the operation of the network described below is based on quantum mechanical principles and is centered on the quantum superposition of the set of paths taken by a single photon as it travels through the network. The states of the photon are physical paths in the network, the lengths of which are determined by a set of delays chosen as in the classical case. Thus, unlike in most quantum computing algorithms devised to date [8], the time variable plays a prominent role in the proposed scheme, with the identification of a specific event at a specific instant in time corresponding to the detection of a Hamiltonian path in the graph. This has the effect of separating the states of the photon over time and allows the detection of said event if and when it occurs. Although no attempt is made at detailing a physical implementation and ideal devices and conditions are assumed, the possibility of an optical arrangement based on currently available devices is hinted at. Such devices appear to be within the limits of precision required to achieve the above-mentioned separation for nontrivial values of N (equation 4.1). The probability of detection of the desired event, however, is very low, and a method to increase it (perhaps on the lines of Grover's inversion about the average [4]) is needed. The possible use of quantum non-demolition (QND) or 'interaction-free' measurements [9, 10] for this purpose is mentioned in the last section.

## 2. A quantum optical system of delays to detect a Hamiltonian path

Let $G = (V_G, E_G)$ be a simple graph of n vertices with vertex set $V_G = \{v_1,...,v_N\}$, edge set $E_G = \{e_{ij} = (v_i,v_j); 1 \leq i \leq N, 1 \leq j \leq N\}$, no self-loops, and adjacency matrix $A = [a_{ij}; 1 \leq i \leq N, 1 \leq j \leq N]$. G is mapped to a system S of optical gratings, channels, and delay lines. A channel is a constrained transmission path with a constant propagation time for all channels, and a delay line is a channel with a fixed delay that can be different for different delay lines. S has two levels.
Level A consists of two stages:
1) an N-slit grating, with the N slits arranged in a circle to prevent bias towards any of the slits; and
2) N channels, with channel j ($1 \leq j \leq N$) directing the grating's output to the input of the j-th 'processor' in row 1 of the matrix described next.
Level B consists of n² 'processing units' arranged in a matrix of n rows and n columns: $U = [u_{ij}; 1 \leq i \leq N, 1 \leq j \leq N]$. Unit $u_{ij}$ ($1 \leq i < N; 1 \leq j \leq N$) feeds its output (as explained next) to unit $u_{i+1\,k}$ ($1 \leq k \leq N$) if and only if $v_j$ is adjacent to $v_k$ in the graph, that is, if and only if $a_{jk} = 1$. It has 3 stages:
1) a delay line that delays inputs from the previous row (or, in the case of row 1, the input from Level A) by a time $\delta_j$ which is distinct for each j;
2) a grating with $k_j$ slits (also arranged in a circle), where $k_j$ is the out-degree of vertex j; $1 \leq k_j < N$; and
3) $k_j$ channels that direct the output of the grating to $k_j$ units in the next row.
(Thus the number of outputs from $u_{ij}$ ($1 \leq i < N$) is the out-degree of vertex j, while the number of inputs to $u_{ij}$ ($1 < i \leq N$) is the in-degree.)
Units in row N have only a delay line followed by a photon detector. A single photon is input to the N-slit grating in Level A.

The feed-forward structure of S causes the network to behave like a non-deterministic Turing machine. There is a one-to-one correspondence between the set of paths taken by the photon (physical paths) through the network and the set of paths through the graph (graph paths). If $d_j$ is the degree of vertex j, then at any unit $u_{ij}$ ($1 \leq i < N, 1 \leq j \leq N$) there are $d_j$ paths to $d_j$ units in the next row. For example, if there is a path $v_3\ v_2\ v_4\ v_2$ in the graph, then there is a physical path for the photon through the units $u_{13}$, $u_{22}$, $u_{34}$, and $u_{42}$, with the photon showing up at the output of $u_{42}$ at time $\delta_3 + \delta_2 + \delta_4 + \delta_2$. It may be physically detected at the output of unit $u_{ij}$ ($1 < i \leq N, 1 \leq j \leq N$) at an arrival time equal to the sum of the delays in a path from row 1 to the output of $u_{ij}$. In arriving at this time, the photon could have taken any of several different paths because there may be more than one distinct path to $u_{ij}$ with the same delay sum along each of the paths. These different physical paths with the same travel time correspond to distinct graph paths that pass through the same number of distinct vertices but in different orders. If no attempt is made to detect the photon at the output of a unit in any of the rows 1 through N-1, it shows up in the output of a single unit $u_{Nk}$ ($1 \leq k \leq N$) in row N where a detector can be used to detect it.

The photon's behavior cannot be described by classical physics as it simultaneously takes all possible paths through the network. It is modeled in quantum mechanical terms that are similar to those used in the theory of the 2-slit (or more generally, k-slit) experiment [11], which requires that any measurement be made only at the output of a unit in row N. If a measurement is made in any of the previous rows, the photon's detection ends any further propagation down the network and the possibility of its tracing a physical path corresponding to a Hamiltonian path in G (if one is present). In the next section the choice of delay values required for the detection of a Hamiltonian path in the graph is discussed.

## 3. Choice of delays

Following [7], let $p_j$ be the j-th prime (actually any N primes can be used). Set the delays to

$$\delta_j = \log p_j; \quad (1 \leq j \leq N) \quad (3.1)$$

With a channel propagation time of zero, the delay sum for a physical path from the input to row 1 to the output of a detector in row N is:

$$\lambda(N) = \sum_{j=1}^{N} c_j \delta_j; \quad 0 \leq c_j \leq \lfloor N/2 \rfloor; \quad \sum_{j=1}^{N} c_j = N \quad (3.2)$$

where the $c_j$'s are non-negative integers. With a non-zero channel propagation time of $\delta_C$, $\lambda$ is increased by $N \delta_C$. (The upper limit on $c_j$ is $\lfloor N/2 \rfloor$ because the graph has no self-loops so that a path in the graph can pass through a vertex no more than $\lfloor N/2 \rfloor$ times.)

If the delays are set as in (3.1), then the delay sum for a non-Hamiltonian path in the network will always differ from that for a Hamiltonian one. Since the delays are real numbers, then, in principle a Hamiltonian path ending in vertex j makes it likely that the photon will appear at the output of $u_{Nj}$ ($1 \leq j \leq N$) in the approximate interval $[\lambda(N), \lambda(N)+\epsilon]$, where $\epsilon$ is some small quantity. If $\epsilon$ is of order $O(N \log_2 N)$ (see equation 4.1 below) the measurement will not confuse this arrival with an arrival at the end of any other path that might have been taken by the photon.

The detection is only 'likely' because it will not occur if the photon happens to take some other path. The choice, though totally random, is not described by conventional probability theory [3]. For the same reason, unlike in the classical behavior of replicated pulses in [7], if no detection occurs it does not mean that the graph is not Hamiltonian. The path itself, when present, may be found by backtracking on the rows and repeating the measurement.

## 4. Minimum delay precision for Hamiltonian path detection

As shown in [7], the smallest difference between the time traveled over a Hamiltonian path and that over a non-Hamiltonian path occurs at the output of row N as the difference between the N-th and (N-1)-st delay. If the first N primes are used in (3.1), this difference $\Delta_{min}$ is given by

$$\Delta_{min} \approx 2/(N \log N) \quad (4.1)$$

(A larger $\Delta_{min}$ results if larger primes are chosen.) This equation holds even with the classical method of [7], but in that architecture there could be an exponential number of physical pulses even half way down the matrix as they make their way to the detectors in the output of row N. Separation of these pulses is required at all levels in the matrix, resulting in a corresponding $\Delta_{min}$ possibly of order $O(2^{-N})$, which means pulses of exponentially small width that are physically unattainable. In contrast, even though there is an exponential number of paths in the quantum scheme there is only one photon following all those paths. Furthermore no attempt is made (or should be made) to detect it along the way. (Even if it were to be done the resolution required to distinguish a partial Hamiltonian path from a non-Hamiltonian one cannot be physically achieved for even moderate values of N.) The detection is done only at the output of each unit in row N, and (4.1) indicates that the resolution required is within practical limits for non-trivial N.

## 5. Detecting and constructing a Hamiltonian path

The procedure to detect a Hamiltonian path in G is as follows. A single photon is input to the N-slit grating preceding the processor matrix at t=0. The N detectors at the output of $u_{Nj}$ ($1 \leq j \leq N$) are set to be open only during the time interval $[\lambda(N), \lambda(N)+\varepsilon]$. If detection occurs at the output of one of the units in row N, it would mean that the graph has a Hamiltonian path. However, as noted previously, failure to detect does not mean that there is no Hamiltonian path.

The procedure to construct a Hamiltonian path is almost identical to that in [7]. The detection procedure is repeated N-1 times over a decreasing number of rows each time, and the path stored as a sequence of vertices. Let the initial detection step indicate that there is a Hamiltonian path ending in vertex j. An expanding one-dimensional vector **h** is used to store the incrementally constructed path and is initialized to [j]. In the first pass the measurement is done with rows 1 through N-1, looking for a pulse at the output of unit $u_{N-1\,k}$, in the time interval $[\lambda(N) - \delta_{N-1\,k}, \lambda(N) - \delta_{N-1\,k} + \varepsilon]$. Such an arrival is 'guaranteed' (in a quantum mechanical sense) for some k, where vertex k is adjacent to vertex j. The path vector is updated to **h** = [k j]. In the second pass the procedure is repeated with rows 1 through n-2 in the time interval $[\lambda(N) - \delta_{N-1\,k} - \delta_{N-2\,m}, \lambda(N) - \delta_{N-1\,k} - \delta_{N-2\,m} + \varepsilon]$, where vertex m is adjacent to vertex k, with j, k, and m distinct. The path vector is now **h** = [m k j]. After the (N-1)-st such pass **h** will contain N distinct vertex indices corresponding to a Hamiltonian path. The time to construct the path is $O(N^2 \log N)$.

## 6. Discussion

In its present form the proposed scheme is only a thought experiment and an idealized one even for the case when a photon is detected only at the output of row N in the expected time interval for a Hamiltonian path. In practice it would require high levels of efficiency of the channels and delay lines used. Whether it can actually be realized cannot be known without constructing a prototype, however small, and performing experiments. Nevertheless an intended implementation could take advantage of advanced optical engineering techniques without requiring special ambient conditions. In particular interferometry works at room temperature, and channels as well as delay lines can be implemented with optical fibers (which, incidentally, provide a high degree of isolation from the environment). Additionally, the ability to generate single photons, which is a basic requirement of the proposed scheme, appears to be closer following recent reports of their successful generation on demand in the laboratory [12].

Should a physical implementation be forthcoming, other improvements are possible. First, the feed-forward matrix can be reduced to a recurrent network with a single row of N units whose delays are set to $\delta_j + \delta$ (where $\delta > \delta_N$) and whose outputs are fed back to the input using the same adjacencies as in the $N \times N$ matrix. Detection of a Hamiltonian path then requires making the measurement in the time interval $[\lambda(N) + (N-1)\delta, \lambda(N) + (N-1)\delta + \varepsilon]$. The procedure to construct the path is obtained by similarly modifying the procedure described above for the feed forward network. Second, instead of a single photon, N photons could be input, one to each of the N units in the first row, leading to the possibility of entangled states that can be operated on to increase the probability of detection [10].

In [7] it is shown that the number of paths is of order $O(4^N)$. As the photon can take any of these paths, the probability of a detector detecting a photon in the expected time interval for a Hamiltonian path is very low. Methods based on unitary transformations may be examined for the possibility of adjusting the probability amplitudes associated with a path in the photon's path space so that a physical path corresponding to a Hamiltonian path (when one exists) is favored over that for a non-Hamiltonian one. One property that might help in making the required distinction is the fact that the travel time for a non-Hamiltonian path as given by the product in equation (3.1) is not square-free in the delay values. There are two techniques that appear likely to prove useful in this regard. The first is the use of quantum nondemolition methods as discussed in [10]. The second is an 'interaction-free' measurement (IFM) that involves 'seeing a photon in the dark' [9, 13]. Thus if it is known that the photon has passed through $u_{ij}$ this knowledge can be used to reduce the probability amplitudes of paths passing through $u_{kj}$, $i+1 < k < N$ (corresponding to paths in the graph that revisit vertex j). Unlike in the optical simulation of quantum logic circuits using single photons [14] where an exponential number of devices matching the exponential number of paths in a circuit is required, the number of IFM devices or QND measurements required in the scheme proposed here is only $O(N^3)$ (even though the number of physical paths of length N-1 is $O(4^N)$).